\documentclass{webofc}
\usepackage[varg]{txfonts}   % Web of Conferences font
\usepackage{listings}
\usepackage{hyperref}
\begin{document}
\title{The {\em Hubble} Catalog of Variables}

\author{\firstname{K.}~\lastname{Sokolovsky}\inst{1,2,3}\fnsep\thanks{\href{mailto:kirx@noa.gr}{\tt kirx@noa.gr}} \and
        \firstname{A.}~\lastname{Bonanos}\inst{1}\fnsep\thanks{\href{mailto:bonanos@astro.noa.gr}{\tt bonanos@astro.noa.gr}} \and
        \firstname{P.}~\lastname{Gavras}\inst{1} \and
        \firstname{M.}~\lastname{Yang}\inst{1} \and 
        \firstname{D.}~\lastname{Hatzidimitriou}\inst{4,1} \and
        \firstname{M.~I.}~\lastname{Moretti}\inst{5,1} \and
        \firstname{A.}~\lastname{Karampelas}\inst{1} \and
        \firstname{I.}~\lastname{Bellas-Velidis}\inst{1} \and
        \firstname{Z.}~\lastname{Spetsieri}\inst{1,4} \and
        \firstname{E.}~\lastname{Pouliasis}\inst{1,4} \and
        \firstname{I.}~\lastname{Georgantopoulos}\inst{1} \and
        \firstname{V.}~\lastname{Charmandaris}\inst{1} \and
        \firstname{K.}~\lastname{Tsinganos}\inst{1} \and
        \firstname{N.}~\lastname{Laskaris}\inst{6} \and
        \firstname{G.}~\lastname{Kakaletris}\inst{6} \and
        \firstname{A.}~\lastname{Nota}\inst{7,8} \and
        \firstname{D.}~\lastname{Lennon}\inst{9} \and
        \firstname{C.}~\lastname{Arviset}\inst{9} \and
        \firstname{B.}~\lastname{Whitmore}\inst{7} \and
        \firstname{T.}~\lastname{Budavari}\inst{10} \and
        \firstname{R.}~\lastname{Downes}\inst{7} \and
        \firstname{S.}~\lastname{Lubow}\inst{7} \and
        \firstname{A.}~\lastname{Rest}\inst{7} \and
        \firstname{L.}~\lastname{Strolger}\inst{7} \and
        \firstname{R.}~\lastname{White}\inst{7}
}

\institute{%1
IAASARS, National Observatory of Athens, Vas.~Pavlou \& I.~Metaxa, 15236~Penteli, Greece 
\and
%2
Sternberg Astronomical Institute, Moscow State University, Universitetskii~pr. 13, 119992 Moscow, Russia
\and
%3
Astro Space Center of Lebedev Physical Institute, Profsoyuznaya Str. 84/32, 117997 Moscow, Russia
\and
%4
%Department of Astrophysics, Astronomy \& Mechanics, 
%Faculty of Physics, University of Athens, 15783 Athens, Greece
Department of Physics, National and Kapodistrian University of Athens, 15771 Ilissia, Greece
\and
%5
INAF-Osservatorio Astronomico di Capodimonte, Salita Moiariello, 16, 80131 Napoli, Italy
\and
%6
Athena Research and Innovation Center, Artemidos 6 \& Epidavrou, 15125~Maroussi, Greece
\and
%7
Space Telescope Science Institute, 3700 San Martin Drive, Baltimore, MD 21218, USA
\and
%8
European Space Agency, Research and Scientific Support Department, Baltimore, MD 21218, USA
\and
%9
European Space Astronomy Centre, Camino bajo del Castillo, Urbanizacion Villafranca del Castillo, Villanueva de la Ca\~{n}ada, 28692 Madrid, Spain
\and
%10
The Johns Hopkins University, Baltimore, MD 21218, USA
          }

\abstract{%
We aim to construct an exceptionally deep ($V \lesssim 27$) catalog of variable objects in
selected Galactic and extragalactic fields visited multiple times by 
the {\em Hubble Space Telescope} (HST).
While HST observations of some of these fields were searched for specific types of variables
before (most notably, the extragalactic Cepheids), we attempt a systematic
study of the population of variable objects of all types at the magnitude 
range not easily accessible with ground-based telescopes.
The variability timescales that can be probed range from hours
to years depending on how often a particular field has been visited.
For source extraction and cross-matching of sources between visits we rely 
on the {\em Hubble} Source Catalog which includes $10^7$ objects detected with 
WFPC2, ACS, and WFC3 HST instruments.
The lightcurves extracted from the HSC are corrected for systematic effects by applying 
local zero-point corrections and are screened for bad measurements.
For each lightcurve we compute variability indices sensitive to a broad
range of variability types. The indices characterize the overall lightcurve 
scatter and smoothness. 
Candidate variables are selected as having 
variability index values significantly higher than expected 
for objects of similar brightness in the given set of observations. 
The {\em Hubble} Catalog of Variables will be released in 2018.
}
\maketitle
%

% Contributed papers: 5 pages; 

\section{Introduction}
\label{sec:intro}

% No need to write why variable stars are important - this was a variable stars meeting.
% Instead - write about the HST.
The {\em Hubble Space Telescope} (HST) is great for deep imaging thanks to the
unique combination of low sky background, sharp point spread function (PSF) and
wide field of view (compared to ground-based adaptive optic systems; \cite{2005aoel.book.....L}). 
It is sensitive to ultraviolet light not accessible from the ground.
% See UVIS Wide-Band (W) Filters at http://www.stsci.edu/hst/wfc3/documents/handbooks/currentIHB/c06_uvis06.html
%
Some fields were imaged by the HST multiple times opening a window to
time-domain studies at faint magnitudes and high spatial resolution. 
High resolution imaging is important to overcome confusion in dense star
fields, like the ones found in nearby galaxies and globular cluster cores.
A number of dedicated variability studies were conducted with the HST
including \cite{2001ApJ...550..554D,2004AJ....127.2738B,2010ApJ...723..737V,2011AJ....141..171J,2013MNRAS.432.3047B,2014MNRAS.442.2381N,2016ApJ...830...10H}.

While the HST has superb internal astrometric precision, obtaining accurate
absolute astrometry has been a challenge in the previous years. 
The positional information provided by the observatory's attitude control system
is limited by the position accuracy of individual 
Guide Star Catalog (\cite{2008AJ....136..735L}) stars.
The {\em Hubble} Source Catalog (HSC, \cite{2016AJ....151..134W}) solves this
problem by cross-matching sources detected in individual HST visits (\cite{2012ApJ...761..188B})
and matching the brightest detected sources to deep reference
catalogs: PanSTARRS, SDSS and 2MASS. The Gaia catalog will be used as reference in future HSC versions.
% References?!?!?!?!?!?

The HSC provides access to a uniform reduction of the majority of publicly
available images obtained with the WFPC2, ACS/WFC, WFC3/UVIS, and WFC3/IR
instruments. The HSC is very inhomogeneous: it includes 112
instrument-filter combinations; some filters are more popular than
others. Most fields were observed only few times and the time between visits varies.
The current HSC version 2 has 89 fields containing $1.7$\,million sources that
were visited at least 5 times while only 7 fields containing $290000$ sources visited $\geq 25$ times.
The HSC is based on visit-combined images that are 
deeper than individual exposures and mostly clean of cosmic rays. 
However, that comes at a price: information about changes on timescales shorter than one visit (that
may last a few hours) is averaged-out and the resulting number of
independent measurements is by a factor of a few smaller than the number of
HST exposures of the field.

We aim to define a set of algorithms that will detect and validate 
candidate variable sources within the HSC, producing the {\em Hubble} Catalog of Variables (HCV).
This is a work in progress that will lead to the release of the first
version of the HCV in 2018. The initial project overview is given by \cite{2017arXiv170300258G}.

\section{Variability search problem}
\label{sec:vi}

It is common in optical photometry that measurement errors are not accurately known.
The contribution of random background variations and photon noise to the
brightness measurements uncertainty can be estimated easily for CCD observations. 
However, the hard-to-quantify residual systematic effects 
(pixel-to-pixel sensitivity variations, charge transfer inefficiency, blending with nearby stars) 
limit the photometric accuracy for brighter objects.
Some measurements get corrupted by cosmic ray hits, CCD cosmetic defects,
incorrect background estimation near frame edges and other undetected 
data processing anomalies. 
Since the majority of field stars are not variable at a few per~cent
level, we utilize them to estimate typical photometric accuracy in a given 
dataset as a function of magnitude.

\begin{figure*}
\centering
\includegraphics[width=0.48\textwidth]{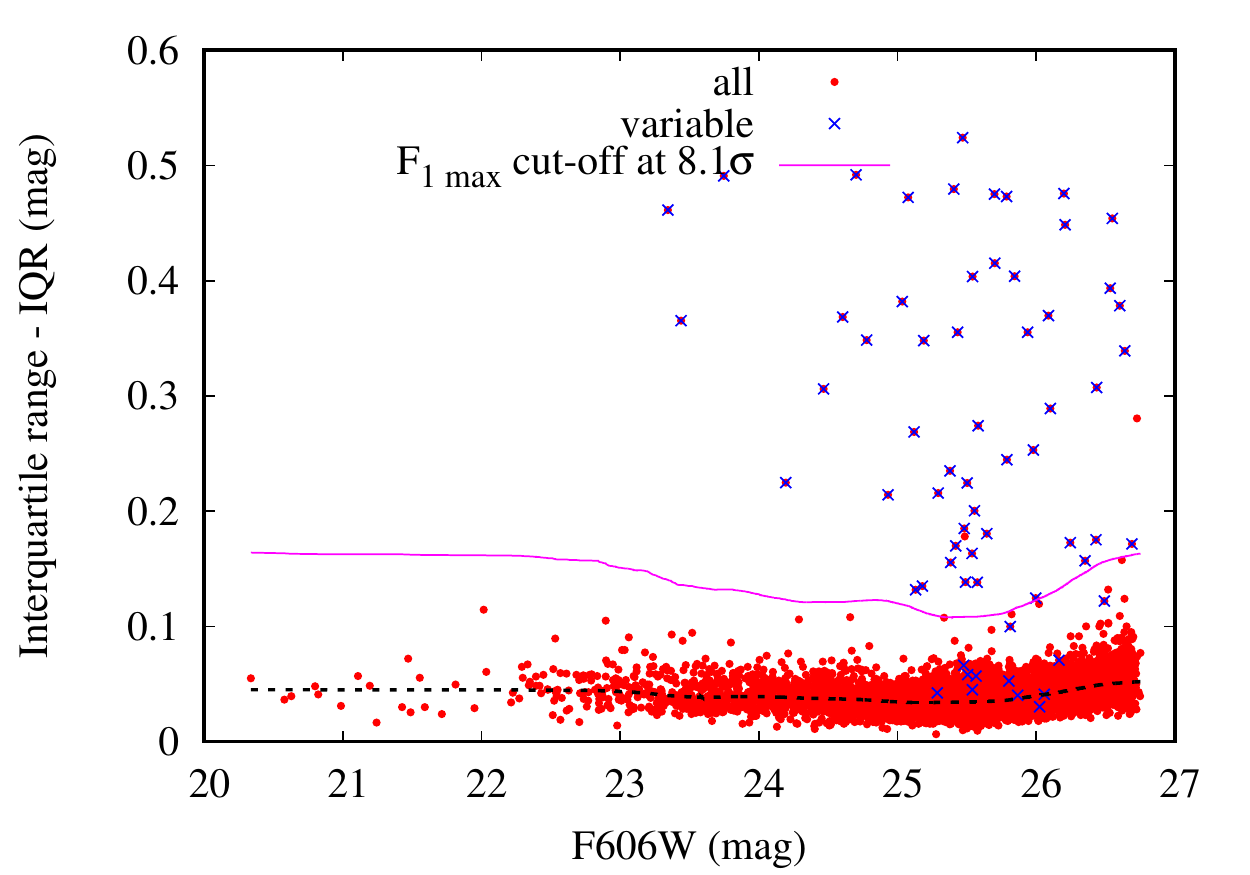}
\includegraphics[width=0.48\textwidth]{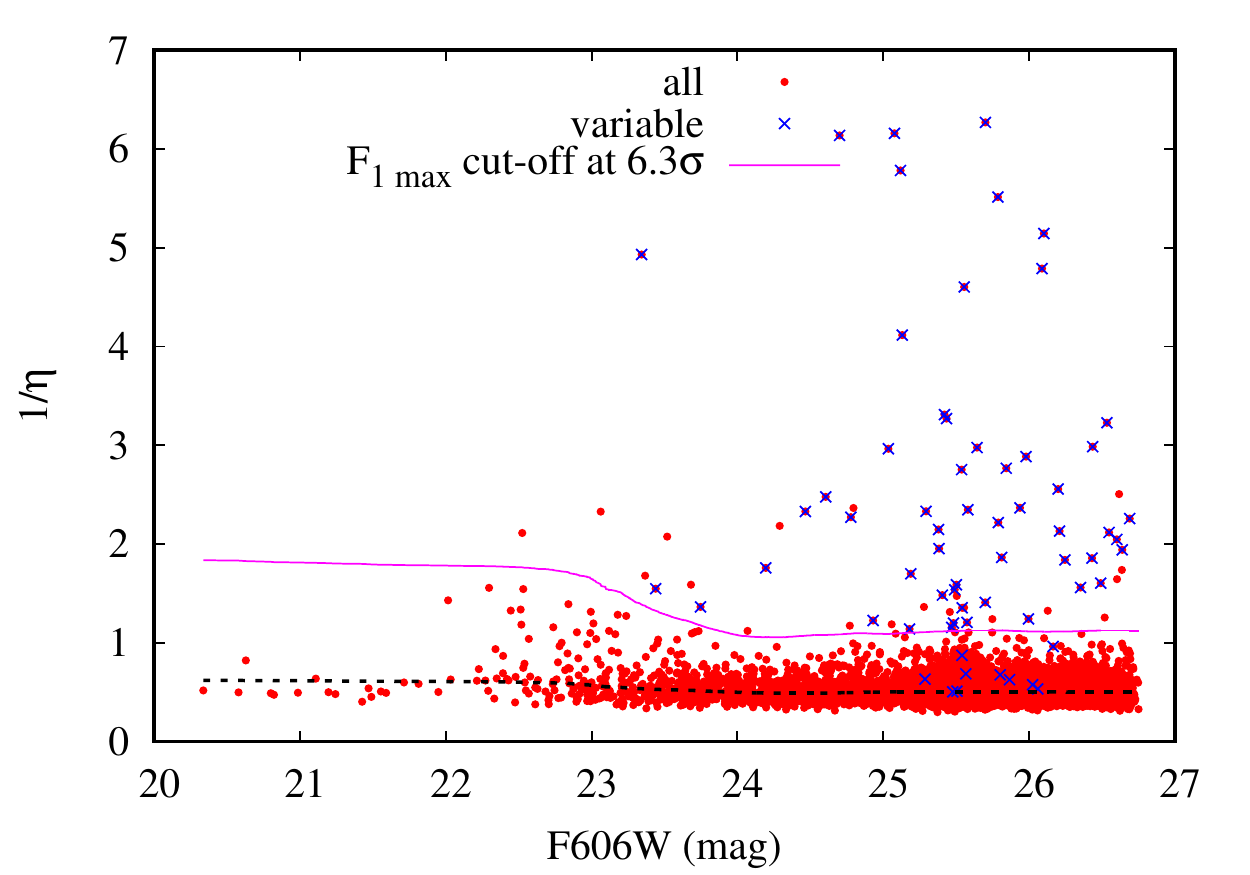}
\caption{Variability indices characterizing scatter (left panel) and
lightcurve smoothness (right panel) plotted as a function of magnitude.
The indices are computed for HSC lightcurves of non-variable objects in
the M31~halo field \cite{2004AJ....127.2738B}. 
Crosses mark the objects in which artificial
non-periodic variability was injected as described in \cite{2017MNRAS.464..274S}. 
Selecting objects above the solid line as candidate variables results in the maximum 
$F_{1}$-score. Dotted line shows the median value of a variability index as a function of
magnitude.}
\label{fig:vi}       % Give a unique label
\end{figure*}

\begin{figure}
\centering
\sidecaption
\includegraphics[width=0.48\textwidth,clip,trim=0cm 0cm 0cm 1cm]{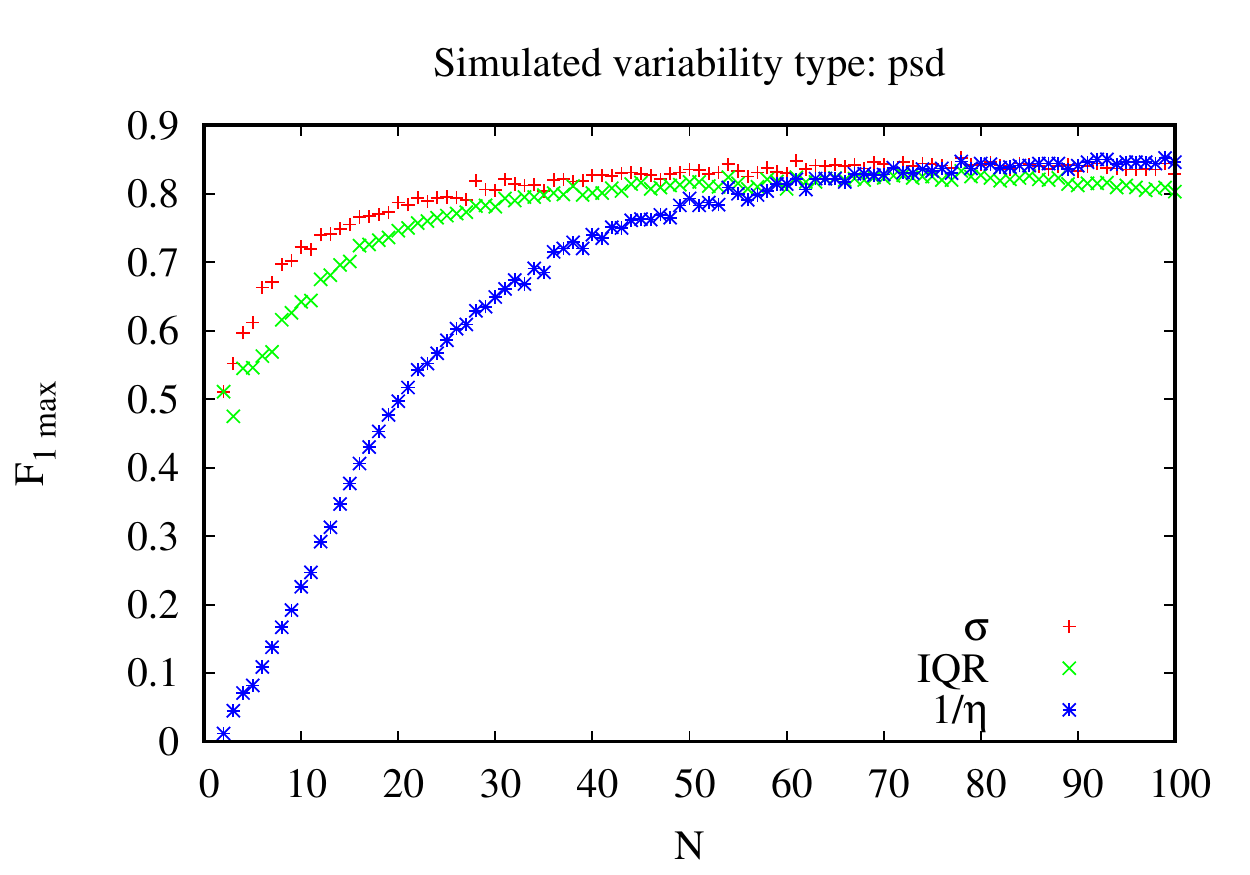}
\caption{The efficiency of variable objects selection characterized by the 
maximum (over all possible cut-off thresholds) $F_{1}$-score as a function of the number of lightcurve
points, $N$, for three variability indices: $\sigma$, IQR and $1/\eta$
(see the references in Table~\ref{tab:varindex}).
Based on simulated non-periodic variability injected in HSC lightcurves (M4 field, \cite{2014MNRAS.442.2381N}).
For the definition of $F_{1}$-score see \url{https://en.wikipedia.org/wiki/F1_score}}
\label{fig:npts}       % Give a unique label
\end{figure}

Ground-based photometry was used by \cite{2017MNRAS.464..274S} to compare 
a number of variability indices that characterize ``how variable'' 
a lightcurve appears by quantifying its scatter and smoothness.
We extend this work by comparing the variability indices listed in Table~\ref{tab:varindex} 
to simulations based on HSC data. We inject variability with random
amplitude into HSC lightcurves of non-variable objects using the technique described 
by \cite{2017MNRAS.464..274S}. The simulations confirm that 
the interquartile range (IQR) of the measured magnitudes,
$m_{i}$ together with the inverse von~Neumann ratio that characterizes
the lightcurve smoothness 
$1/\eta = { \sum\limits_{i=1}^N(m_i-\bar{m})^2}/{\sum\limits_{i=1}^{N-1}(m_{i+1} - m_{i})^2}$
(where $\bar{m}$ is the mean magnitude and $N$ is the number of measurements)
can recover a broad range of variability patterns 
and are robust against individual outlier measurements.
These indices do not depend on the estimated errors (Table~\ref{tab:varindex}) 
that may be unreliable. Figure~\ref{fig:vi} presents IQR and $1/\eta$
indices as a function of magnitude in one of the simulations. 
Figure~\ref{fig:npts} illustrates how the variability detection
efficiency of the indices changes as a function of the number of points in a lightcurve.

\begin{table}
\centering
    \caption{Variability indices computed by the HCV pipeline.}
    \label{tab:varindex}
    \begin{tabular}{r@{~~}c@{~~}l r@{~~}c@{~~}l}
    \hline%\hline
Index                                            &   Errors  & Ref.                          & Index                                           &  Errors  & Ref. \\
    \hline
\multicolumn{3}{c}{\it Indices quantifying lightcurve scatter}                                       & time-weighted Stetson's~$J_{\rm time}$          & $\checkmark$ & \text{\cite{2012AJ....143..140F}}  \\
reduced $\chi^2$ statistic -- $\chi_{\rm red}^2$ & $\checkmark$ & \text{\cite{2010AJ....139.1269D}}  & clipped Stetson's~$J_{\rm clip}$                & $\checkmark$ & \text{\cite{2017MNRAS.464..274S}}  \\
weighted std. deviation -- $\sigma$              & $\checkmark$ & \text{\cite{2008AcA....58..279K}}  & Stetson's~$L$ index                             & $\checkmark$ & \text{\cite{1996PASP..108..851S}}  \\
median abs. deviation -- ${\rm MAD}$             &              & \text{\cite{2016PASP..128c5001Z}}  & time-weighted Stetson's~$L_{\rm time}$          & $\checkmark$ & \text{\cite{2012AJ....143..140F}}  \\
interquartile range -- ${\rm IQR}$               &              & \text{\cite{2017MNRAS.464..274S}}  & clipped Stetson's~$L_{\rm clip}$                & $\checkmark$ & \text{\cite{2017MNRAS.464..274S}}  \\
robust median stat. -- ${\rm RoMS}$              & $\checkmark$ & \text{\cite{2007AJ....134.2067R}}  & consec. same-sign dev. -- {\it Con.}            &              & \text{\cite{2000AcA....50..421W}}  \\
norm. excess variance -- $\sigma_{\rm NXS}^2$    & $\checkmark$ & \text{\cite{1997ApJ...476...70N}}  & excursions -- $E_x$                             & $\checkmark$ & \text{\cite{2014ApJS..211....3P}}  \\
norm. peak-to-peak amp. -- $v$                   & $\checkmark$ & \text{\cite{2009AN....330..199S}}  & autocorrelation -- $l_1$                        &              & \text{\cite{2011ASPC..442..447K}}  \\
\multicolumn{3}{c}{\it Indices quantifying lightcurve smoothness}                                    & inv. von~Neumann ratio -- $1/\eta$              &              & \text{\cite{2009MNRAS.400.1897S}}  \\
Welch-Stetson index -- $I$                       & $\checkmark$ & \text{\cite{1993AJ....105.1813W}}  & excess Abbe value -- $\mathcal{E}_\mathcal{A}$  &              & \text{\cite{2014A&A...568A..78M}}  \\
Stetson's~$J$ index                              & $\checkmark$ & \text{\cite{1996PASP..108..851S}}  & $S_B$ statistic                                 & $\checkmark$ & \text{\cite{2013A&A...556A..20F}}  \\
    \hline
    \end{tabular}
\end{table}

%
%For tables, please use the syntax of Table~\ref{tab:tab-1}.
%
%\begin{table}
%\centering
%\caption{Please write your table header here.}
%% For LaTeX tables you can use
%\begin{tabular}{lll}
%\hline
%first & second & third  \\\hline
%number & number & number \\\hline
%\end{tabular}
%\label{tab:tab-1}       % Give a unique label
%\end{table}
%%

\section{HCV data pre-processing and variability detection pipeline}
\label{sec:preproc}

\begin{figure*}
\centering
\includegraphics[height=5cm,clip,trim=1.0cm 0cm 1.0cm 0.35cm]{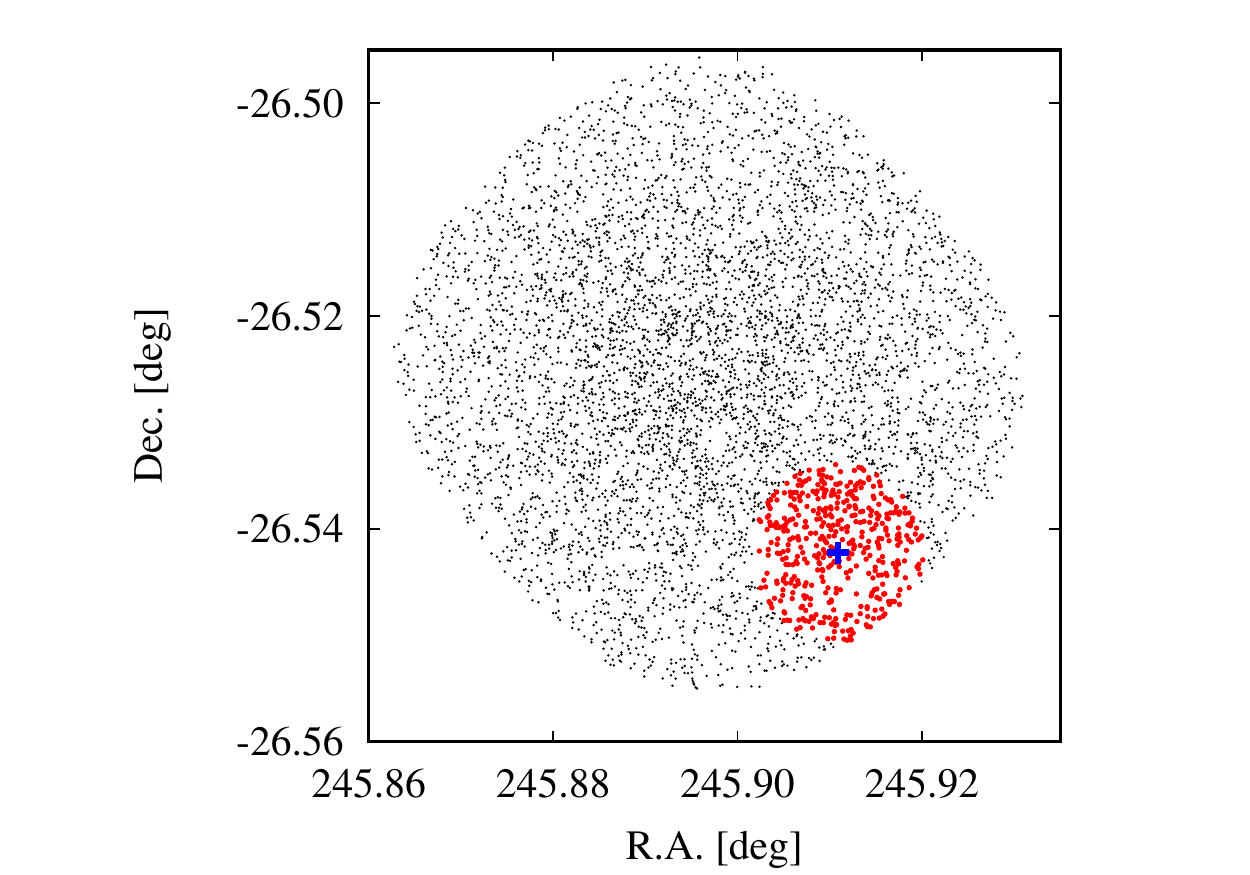}
\includegraphics[height=5cm,clip,trim=0cm 0cm 0cm 1.0cm]{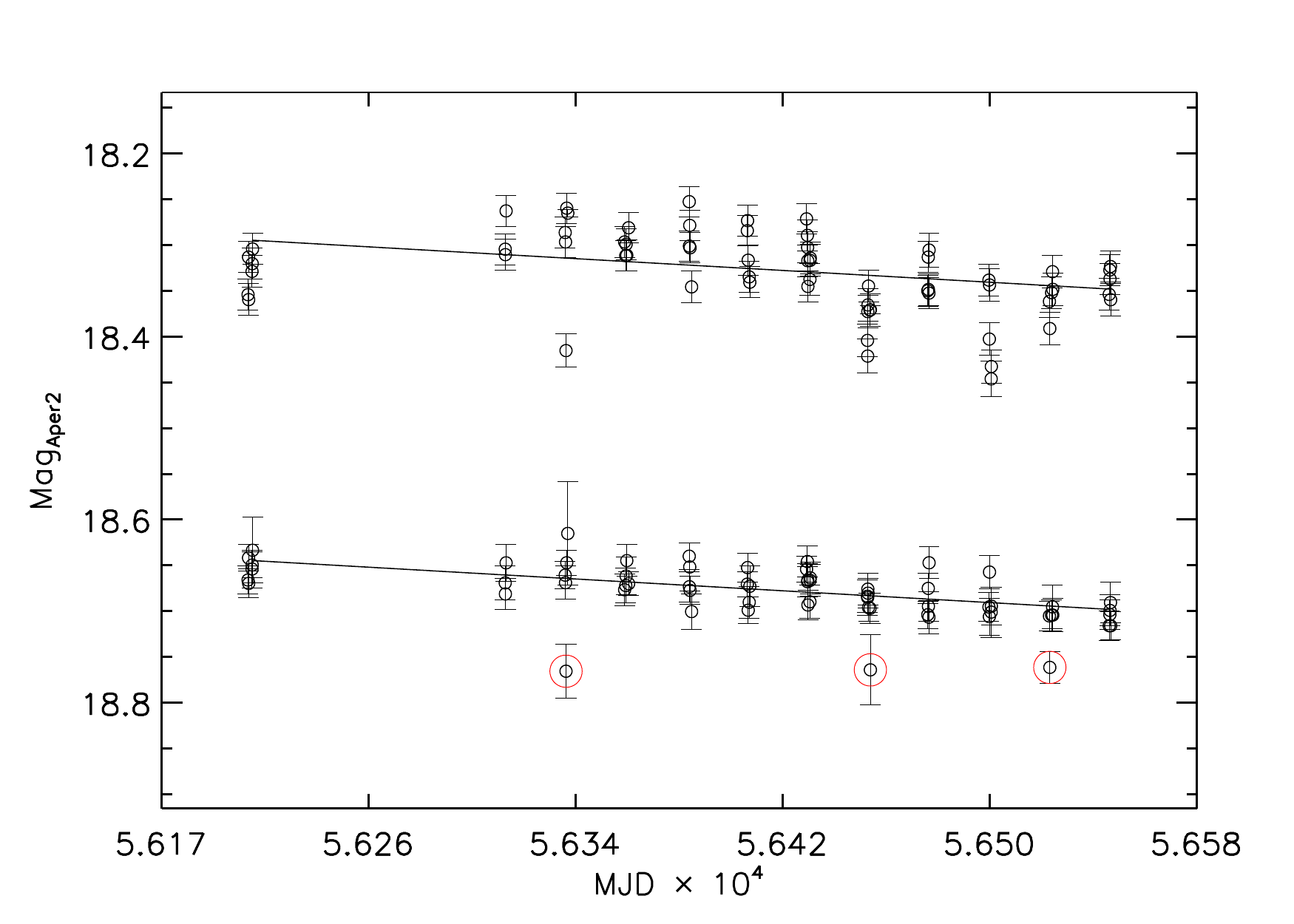}
\caption{Lightcurve pre-processing. Left: spatial distribution of HSC
objects in the M4 field \cite{2014MNRAS.442.2381N}. Highlighted are the objects used to determine local
magnitude zero-point correction for the object marked with the cross. Right:
lightcurve of that object before (top) and after (bottom) applying the local
correction. The circles mark identified outlier points. The two
lightcurves are shifted along the magnitude axis for clarity.}
\label{fig:lc}       % Give a unique label
\end{figure*}

The HSC data are imported and grouped according to the observed field.
The measurements of objects near frame edges or obtained with an old
version of the image processing pipeline are excluded as unreliable.
%(such data filtering might not be necessary in future versions of HSC). 
For each object all its measurements in a given filter that pass the
above selection are collected to construct a lightcurve.
Each lightcurve is fitted with a straight line using robust regression
and outliers from the fit are identified. If a high percentage of
measurements obtained during some visit are identified as outliers in
the corresponding lightcurves, all measurements associated with this visit
are discarded (bad image). For each of the remaining visits, for each object
a local zero-point correction is computed as the mean difference between
the measured magnitudes and the ones predicted by the robust line fit for
all objects within a specified radius around the object being corrected
(Fig.~\ref{fig:lc}). This should compensate for the residual large-scale 
sensitivity variations across the image.

A set of corrected HSC lightcurves of objects observed in a given field with the same 
instrument-filter combination is the basic unit of variability search.
For each lightcurve in a set we compute variability indices and select
as candidate variables the objects that have variability index values 
significantly higher than the typical value for their brightness. 
Magnitude-dependent cuts in robust indices IQR and $1/\eta$ are currently used to select candidate variables
(Fig.~\ref{fig:vi}). We are investigating ways
to efficiently combine information captured by all indices listed in 
Table~\ref{tab:varindex} by means of the principal component analysis and machine learning.

The HCV pipeline is implemented in {\scshape Java} and parallelized using 
the {\scshape Apache Spark} framework. The variability indices implementation 
in the pipeline is consistent with their implementation in {\scshape VaST}
(\cite{2017arXiv170207715S}) which is also used for lightcurve visualization and testing
while a dedicated HCV visualization software is being developed.

\section{Summary}
\label{sec:summ}

\begin{itemize}
\item The HCV is a catalog of variable objects derived from the HSC. It will be released in 2018.
\item The catalog is very heterogeneous due to the nature of the HSC dataset.
It covers selected fields in the Galaxy, the Local Group and beyond.
\item The HCV is very deep, it ventures into poorly explored region of variability parameter space.
\item Data pre-processing and variability detection techniques used for HCV are applicable to other
multi-epoch surveys.
\end{itemize}

\begin{acknowledgement} 
\noindent\vskip 0.2cm
\noindent {\em Acknowledgments}: This work is supported by ESA under contract No.~4000112940.
\end{acknowledgement}

% BibTeX or Biber users please use (the style is already called in the class, ensure that the "woc.bst" style is in your local directory)
% \bibliography{name or your bibliography database}

\end{document}